# Laser-induced synthesis and decay of Tritium under exposure of solid targets in heavy water

E.V. Barmina[1], P.G. Kuzmin[1], S.F. Timashev[2,3], and G.A. Shafeev[1]

[1] Wave Research Center of A.M. Prokhorov General Physics Institute of the Russian Academy of Sciences, Moscow, Russian Federation

[2] L.Ya. Karpov Institute of Physical Chemistry, Moscow, Russian Federation

[3] National Research Nuclear University MEPhI, Moscow, Russian Federation

Abstract

The processes of laser-assisted synthesis of Tritium nuclei and their laser-induced decay in cold plasma in the vicinity of solid targets (Au, Ti, Se, etc.) immersed into heavy water are experimentally realized at peak laser intensity of $10^{10}$-$10^{13}$ W/cm$^2$. Initial stages of Tritium synthesis and their laser-induced beta-decay are interpreted on the basis of non-elastic interaction of plasma electrons having kinetic energy of 5-10 eV with nuclei of Deuterium and Tritium, respectively.

Recent theoretical work shows the capability of laser radiation to directly excite nuclear levels of energy [1]. However, exciting a nuclear transition would require an X-ray or gamma-ray lasers with an intensity greater than $10^{20}$ Watts/cm$^2$. At the same time, laser generated photons can indirectly influence the nuclear levels and alter or induce nuclear reactions. Laser beams at intensity level of $10^{18}$ W/cm$^2$ are capable of inducing nuclear transformations under laser exposure of solid targets in vacuum. This can be achieved at pico- and femtosecond laser pulse durations [2-4]. The emission of alpha-particles was observed under exposure of $^{11}$B target to picosecond laser pulses. Experiments were carried out on the "Neodymium" laser facility at the pulse energy of 10–12 J and pulse duration of 1.5 ps. The composite targets $^{11}$B + {CH$_{2n}$} placed in vacuum were used. The yield of $10^3$ alpha particles per pulse has been detected at peak laser intensity on the target of $2 \times 10^{18}$ W/cm$^2$ [5]. In accordance with [5], the mechanism of initiation of a neutron-less fusion reaction is due to the acceleration of electrons and protons in the plasma separated by strong laser field.

New possibilities of laser initiation of nuclear reactions have been demonstrated at peak laser intensity levels of $10^{10}$-$10^{13}$ W/cm$^2$ [6-9]. This approach is based on the laser exposure of nanoparticles suspended in a liquid (colloidal solution). The crucial role in laser-induced acceleration of nuclear decays belongs to nanoparticles (NPs) and nanostructures (NS) on the target immersed into liquid that interact with laser beam in presence of unstable isotopes. In fact, the possibility of initiation of nuclear transformations in cold laser-induced plasma with electron temperature of 5-10 eV has been shown. Each nanoparticle can be considered as a target for laser exposure. Unlike bulk targets, metal nanoparticles are optically thin. The critical density is reached from the onset of laser pulse; this is the density of free electrons in the nanoparticle but owing to the small size of the target the reflection is negligible. In this case the whole nanoparticle is heated up by the laser beam homogeneously. The same applies to the late stages of laser heating – the material of the target is ionized and expands in the shape of spherical cavity but the laser radiation is still absorbed by the plasma due to the small size of the cavity.

Acceleration of decay induced by laser exposure at intensity level of $10^{13}$ W/cm$^2$ has been reported for aqueous solutions of $^{232}$Th and $^{137}$Cs impurity [6]. In the latter case the beta-decay accelerated by laser exposure is demonstrated. The results of laser exposure are very different in H$_2$O and D$_2$O. Later it has been experimentally demonstrated that laser exposure of metallic targets (Au, Ag, Pd, Be) in aqueous solutions of $^{238}$UO$_2$Cl$_2$ leads to generation of nanoparticles of corresponding materials in the bulk of the liquid and simultaneously shifts the secular equilibrium in the activity of nuclides of its branching, such as $^{234}$Th and $^{234m}$Pa [7-9]. It is pertinent to note that laser exposure affects both the alpha- and beta-decays, and the effect of acceleration is not sensitive to the type of water used. The variation of activity of nuclides induced by laser exposure demonstrates isotopic selectivity: the activity of $^{234}$Th and $^{234m}$Pa decreases after laser exposure, while the activity of $^{235}$U remains at the initial level. This result is in good agreement with previously observed activity variation of Uranium under electro-explosion of metallic foils in aqueous solutions of Uranium salt [10]. Another example of laser-induced nuclear reactions is the observation of transmutation of stable nuclei $^{196}$Hg – $^{197}$Au induced by laser exposure of $^{196}$Hg nano-drops in D$_2$O [11].

In the last 20 years, especially after the first works on the possibility of realization of thermonuclear reactions via the electrolysis of heavy water with a Pd cathode, during which excess heat is evolved and neutrons and tritium nuclei are detected [12, 13], the number of problems that arise has increased. If the results from works on "cold fusion" raised (and still raise) certain doubts due to their irreproducibility in different laboratories, experimental evidence on the possibility of the transmutation of various nuclei upon the laser excitation of nanoparticles [6-9, 11] and electro-explosions of metallic electrodes in aqueous solutions of salts of unstable isotopes [10] are well reproducible. These results confirm the possibility of initiation of nuclear transformations in cold plasma.

Tritium is another unstable isotope with half-life of 12.2 years. It would be interesting to induce its beta-decay by exposure to laser radiation, as it was done previously for other nuclides. In this work we demonstrate for the first time the nuclear synthesis and decay of Tritium under laser exposure of various targets in $D_2O$.

Different targets were used for exposure of bulk metal targets in a heavy water with some tritium content ($D_2O + \xi DOT$), where $\xi$ stands for molar fraction of $D_2O$ with T. The content of T was measured according to its beta-activity with the use of a beta-spectrometer with fluorescent dye. The accuracy of measurements was 1%. Deuterium content was 98.8 %. Two laser sources were used for exposure of various targets in heavy water. The first source was a copper vapor laser, pulse duration of 15 ns, wavelength of 510.6 nm, and repetition rate of 15 kHz. The estimated peak power on the target was $10^{10}$ W/cm$^2$. The second laser source was a Nd:YAG laser with pulse duration of 10 ps and repetition rate of 50 kHz either at first (1064 nm) or at its second harmonics (532 nm). The estimated peak power on the target was $10^{13}$ W/cm$^2$ and $5 \times 10^{11}$ W/cm$^2$ at 1064 and 532 nm, respectively.

Laser exposure of various targets was carried out by focusing laser beam on the target with spherical lens. Typically, 2 ml of water was placed inside a glass cell cooled by flowing water. In another set of experiments the metallic target was cathodically biased with respect to another electrode made either of stainless steel or Platinum wire 0.5 mm in diameter. The applied voltage was 25 V. To ensure the conductivity of the solution, 15 mg of metallic Na were added to 5 ml of $D_2O$ thus producing conducting solution of NaOD.

In the first set of experiments the ablation of targets was carried out in pure $D_2O$ (supplied by Euriso-top) without Tritium content. In this case the content of DOT molecules in $D_2O$ was at the background level of the beta-spectrometer, which corresponds to $\xi = 3.54 \cdot 10^{-14}$. In one set of experiments the voltage was not applied to the target, in the second set of experiments the ablation was carried out simultaneously with electrolysis. Application of voltage to electrodes is accompanied by reduction of Deuterium (and Tritium) on the target as bubbles.

The results are presented in Table 1. Laser ablation of targets without electrolysis leads to the increase of Tritium activity to the level which is about 20 times above the background ($10^{-7}$ Curie/l). The synthesis of Tritium is even more pronounced if the ablated target is cathodically shifted. Ablation of either Ti or Au targets with electrolysis results in synthesis of Tritium to the level, which is $10^3$ times above the background.

Table 1. Laser exposure of targets in pure $D_2O$ ($\xi = 3.54 \cdot 10^{-14}$)

| № | Material | Kind of liquid | Wavelength of laser beam, nm | Intensity of laser radiation, W/cm² and pulse width | Time of irradiation, min | Electrolysis | Activity, Ci/l | Change with respect to initial level |
|---|---|---|---|---|---|---|---|---|
| 1 | Initial $D_2O$ | $D_2O$ | - | - | - | - | $8.64 \times 10^{-8}$ $\xi=3.54 \cdot 10^{-14}$ | - |
| 2 | Au | NaOD | 532 | $5 \times 10^{11}$ 10 ps | 60 | + | $4.06 \times 10^{-4}$ $\xi=1.66 \cdot 10^{-10}$ | Increase $4.7 \times 10^3$ |
| 3 | Au | NaOD | 1064 | $10^{13}$ 10 ps | 32 | - | $1.62 \times 10^{-6}$ $\xi=6.63 \cdot 10^{-13}$ | Increase by factor 18.8 |
| 4 | Ti | NaOD | 532 | $10^{11}$ | 62 | + | $4.74 \times 10^{-4}$ $\xi=1.94 \cdot 10^{-10}$ | Increase $5.5 \times 10^3$ |
| 5 | Ti | NaOD | 1064 | $10^{13}$ | 90 | - | $1.57 \times 10^{-6}$ $\xi=6.42 \cdot 10^{-13}$ | Increase by factor 18.2 |

The following set of experiments was carried out in heavy water with initial Tritium content. Laser ablation of bulk targets for 40-60 min leads to generation of nanoparticles (NPs). In case of Au and Se targets these NPs are made of Au and Se, respectively [15, 16]. Both these colloids absorb in the visible range of spectrum and are reddish in appearance. The generated colloids are unstable towards sedimentation, and NPs can easily be separated from the solution. The solution without NPs (or with small remaining fraction of NPs) was tested for Tritium content about one week after laser exposure. Tritium content in the initial $D_2O$ was also measured in the same set of samples. Then in another set of experiments the generated colloids were exposed to laser radiation in absence of the target.

The results of laser exposure of bulk Au and Se targets on T content are presented in Table 2.

Table 2. Laser exposure of Se and Au targets in $D_2O$ with DOT without electrolysis.

| № | Material | Wavelength of laser beam, nm | Intensity of laser radiation, W/cm² | Time of irradiation, min | Activity, Ci/l |
|---|---|---|---|---|---|
| 1 | Initial $D_2O$+ DOT | - | - | - | $2.96 \times 10^{-5}$ ($\xi = 1.21 \cdot 10^{-11}$) |
| 2 | Se | 510 (Cu laser, 15 ns) | $10^{10}$ | 60 | $4.43 \times 10^{-6}$ Decrease by a factor 6.7 |
| 3 | Se | 532 (10 ps) | $5 \times 10^{11}$ | 40 | $5.69 \times 10^{-5}$ Increase by a factor 1.9 |
| 4 | Au | 510 (Cu laser, 15 ns) | $10^{10}$ | 60 | $2.87 \times 10^{-5}$ Small changes |
| 5 | Au | 532 (10 ps) | $5 \times 10^{11}$ | 60 | $5.57 \times 10^{-5}$ Increase by a factor 1.9 |

Tritium content increases by a factor of 1.9 after ablation of a bulk Se target with 10 ps pulses of the second harmonics of a Nd:YAG laser. Further laser exposure of Se NPs solution does not alter this increased level of Tritium activity. Again, exposure of the colloidal solution to radiation of a Cu vapor laser leads to decrease of T activity by a factor of **6.7**. Qualitatively similar results are obtained for NPs of Au (Table 3).One can see that activity of Tritium (its content) is higher by a factor of 1.8 than in the initial $D_2O$ right after laser generation of Au NPs with picosecond laser pulses. Further laser exposure of the colloidal solution alone does not alter much the activity of T. On the contrary, small decrease of T activity is observed in case of exposure of Se colloidal solution with 15 ns pulses of a Cu vapor laser.

The exposure of different targets (Au, Ag, Be) to radiation of a 15 ns laser pulses has small effect on the activity of Tritium. The exception is exposure of Se target to this radiation that markedly reduces T content in the exposed solution (see Table 2).

Laser exposure of targets that are cathodically biased in $D_2O$ with relatively high DOT content leads to some decrease of Tritium activity. These results are presented in Table 3.

Table 3. Decay of Tritium activity during laser exposure of different targets with electrolysis.

| No | Material | Kind of liquid | Wavelength of laser beam, nm | Intensity of laser radiation, W/cm$^2$ | Time of irradiation, min | Activity, Ci/l |
|---|---|---|---|---|---|---|
| 1 | Initial $D_2O$+ DOT | $D_2O$+ DOT | - | - | - | $1.22 \times 10^{-3}$ ($\xi = 0.5 \cdot 10^{-9}$) |
| 2 | Au | NaOD | 532 | $10^{11}$ | 61 | $1.12 \times 10^{-3}$ |
| 3 | Au | NaOD | 1064 | $10^{13}$ | 30 | $9.03 \times 10^{-4}$ |
| 4 | Ti | NaOD | 532 | $10^{11}$ | 60 | $1.11 \times 10^{-3}$ |
| 5 | Ti | NaOD | 1064 | $10^{13}$ | 90 | $3.35 \times 10^{-4}$ |
| 6 | Pd | NaOD | 1064 | $10^{13}$ | 90 | $1.22 \times 10^{-3}$ |

The final value of activity after laser exposure is around $10^{-3}$ Ci/l. The variation of activity caused by laser ablation is well pronounced even at high level of initial Tritium activity. As it follows from data presented in Table 1, at low initial Tritium content the increase of Tritium content after laser exposure is observed. One may suggest that this result is the competition between laser-induced decay of Tritium and its formation. Unlike ablation of Au in $D_2O$, NPs generated from Ti target are not metallic. These NPs have bluish color and consists presumably of oxides or/and hydroxides of Titanium. Au NPs generated by ablation of a bulk Au target with electrolysis are unusually blue in appearance, which corresponds to elongated Au NPs. Under long laser exposure the blue color transforms into grey, and maximum of absorption shifts to near infrared region of spectrum.

The efficiency of nuclear processes occurring during the course of heavy water electrolysis can depend on the character of roughness of the electrode surfaces on a nanometer scale, the "spikiness" parameters [17, 18] in particular. Indeed, it is precisely in the regions of the sharpest surface relief alterations that high electric field strengths making for the acceleration of electrons and high mechanical stresses depressing the activation barriers for electrochemical processes can both get realized. This parameter is out of control in most experiments with electrolysis of heavy water. On the contrary, laser ablation of metallic targets by sub-nanosecond laser pulses leads to formation of self-organized nanostructures (NS) on the target. The average size and density of NS depends on laser fluence on the target and target material. Typical view of such NS on Ti and Au target ablated in water with 10 ps laser pulses are presented in Fig. 1.

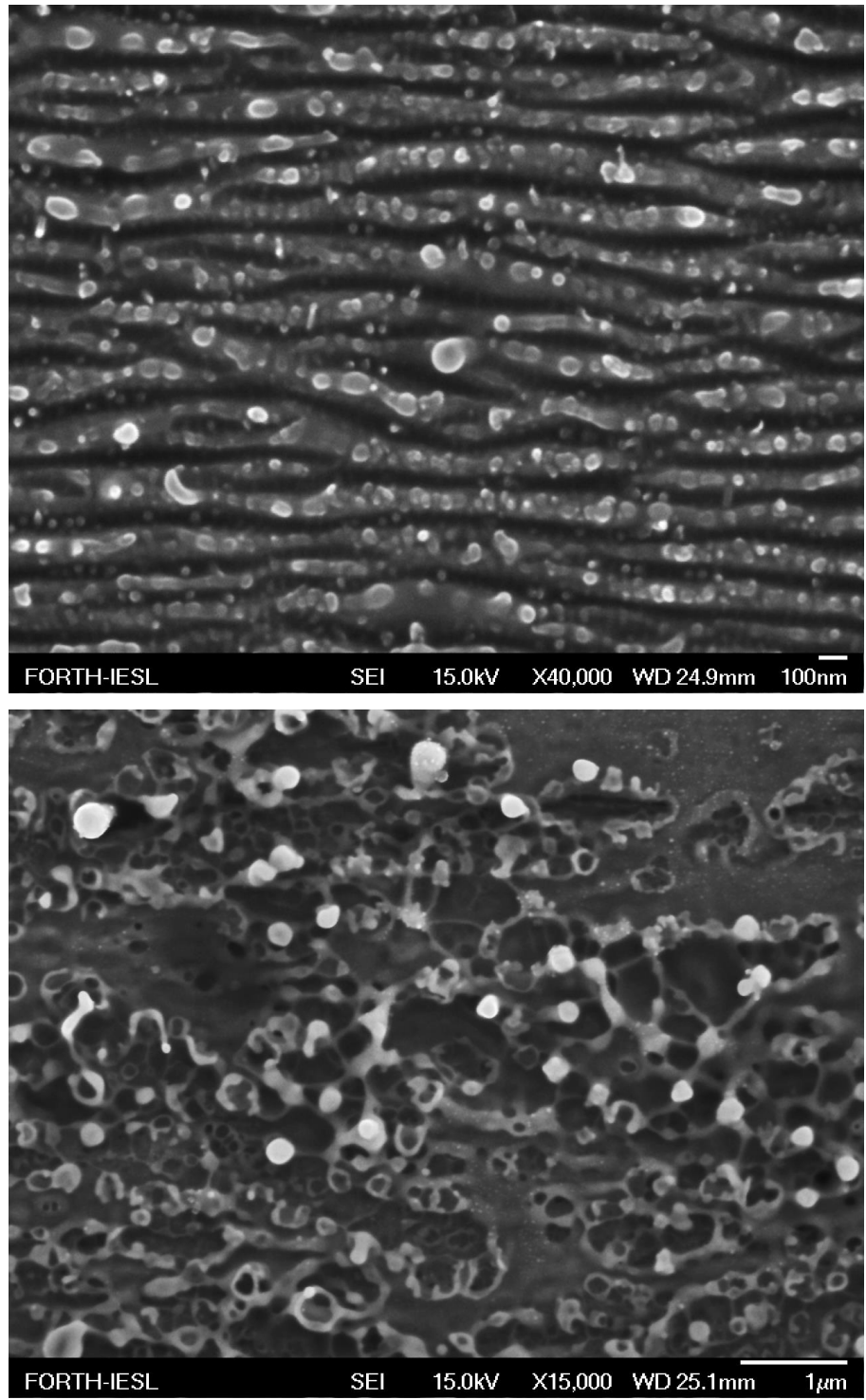

Fig. 1. Typical view of nanostructures generated by laser ablation of Ti (a) and Au (b) targets by the second harmonics of a Nd:YAG laser with pulse duration of 10 ps. Protruding structures look brighter. Scanning Electron Microscope with Field Emission. Scale bar denotes 100 nm (a) and 1 μm (b).

NS on Au generated by laser ablation in water show high activity in Surface Enhanced Raman Scattering with amplification factor of $10^4$-$10^8$ [19, 20]. Therefore, laser ablation technique used in the present work provides guaranteed "spikiness" of the target that may generate high electric fields in the vicinity of the ablated target.

The observed phenomena of T synthesis cannot be ascribed to nuclear fusion in its classical sense. There should be a low-energy channel that may alter the interaction of deuterons. One may suggest the following scheme of reactions [21, 22]. If the original nucleus is nonradioactive (for the sake of definiteness, let it be a deuteron $d^+$) and the kinetic energy $E_e$ of the electron is too low for any nuclear process involving the breakdown of the original nucleus to take place, the electron-nucleus interaction can result in the emission of a neutrino ν (an irreversible process!) and formation in the nucleus of an intermediate vector boson $W^-$ as a virtual particle (the electron cannot directly interact with a quark). Thereafter there forms a d quark, as a result of interaction between the virtual vector boson $W^-$ and one of the u quarks of a nuclear proton, and the latter converts into a virtual neutron. No actual neutron can form in the nucleus. True, if this were the case, the nucleus being formed would decay, because a two-neutron system is unstable [23]. But for such decay into two neutrons to take place, the mass of the nucleus being formed, which is equal to the sum of the masses of the deuteron and electron, is insufficient. (In the case under consideration, the energy deficit $\Delta Q$ is $(m_p + m_e - 2m_n)c^2 \approx -3.01$ MeV, where $c$ is the velocity of light in vacuum and $m_p$, $m_e$, and $m_n$ are the masses of the proton, electron, and neutron, respectively.) The latter means that with the formation of a second neutron in the nucleus being impossible, the state of the nuclear matter here cannot be represented in the standard way, i.e., in the form of combination of a definite number (two in the given case) nucleons, and described in terms of the well-known nucleon-nucleon interaction. For this reason, the conclusions drawn in [23] that the existence of a nuclear stable dineutron is impossible does not relate to the nucleus being introduced here with its mass less than the total mass of two neutrons, despite its zero electric charge, baryon number equal to 2, and zero leptonic charge. Actually one should speak in this case about the origin of a non-standard, "non-balanced" or "inner shake-up" ("in-shake-up") [21, 22] state of nuclear matter. Obviously, the relaxation of such an "inner shake-up" ("in-shake-up") state of the nucleus, which owes to the aforementioned imbalance of nuclear matter, must end with its $β$-decay – the emission of a free electron $e^-$ and an antineutrino $\tilde{\nu}$ – accompanied by the formation of a deuteron. For the sake of definiteness, therefore, we will refer to such a nucleus as $β$-dineutron and represent the sequence of the processes being discussed as follows:

$$e^-_{he} + d^+ \rightarrow {}^2n_{isu} + \tilde{\nu}, \qquad (1)$$

$$^2n_{isu} \rightarrow d^+ + e^- + \tilde{\nu}. \qquad (2)$$

In these expressions, the subscript "he" on the notation of the initial electron points to the initiative character of interaction between the electron and deuteron in the formation of the β-dineutron $^2n_{isu}$ in the in-shake-up state (the latter being indicated by the subscript "isu"). Inasmuch as the threshold of such an "inelastic scattering" of the electron by the deuteron is solely determined by the doubled rest mass of the neutrino and amounts to less than 1 eV, the electron kinetic energies of $E_e \sim 10$ eV will apparently be adequate for one to observe such processes. This means that the mass of the β-dineutron practically coincides, as noted above, with the mass of the deuterium atom. It should also be noted that at low energies actually the entire internal "organization" of the processes involving quarks and vector bosons, qualitatively represented above, can only be reflected in the magnitude of the effective gross-interaction constant. In that case, the characteristic half-life of the β-dineutron determined by the weak nuclear interactions can be long enough. The latter means the Triton in the experimental conditions of present work may be synthesized according to the following equation:

$$d^+ + {}^2n_{isu} \rightarrow t^+ + n + Q(3.25 MeV). \qquad (3)$$

In turn, Tritium nucleus formed in the cold plasma may also interact with electrons. According to [21,22], decay of Tritium may be accelerated ("e-catalysis") owing to this interaction and formation of an intermediate product β-trineutron $^3n_{isu}$:

$$t^+ + e^- \rightarrow {}^3_0 n_{isu} + \nu \rightarrow {}^3_2 He^{2+} + 2e^- + \nu + 2\tilde{\nu} + Q(0.019 MeV), \qquad (4)$$

All experimental results presented above can be interpreted on the basis of the processes (1), (3), and (4). Indeed, processes (1)-(3) dominate at low initial Tritium content in $D_2O$ ($10^{-7}$ Curie/l) and presence of electrons having kinetic energy $E_e \sim 5$-$10$ eV in cold plasma. As a result, Tritium content increases by 3 orders of magnitude. On the other hand, predominant β-decay of Triton takes place under relatively high initial Tritium beta-activity $\sim 10^{-3}$ Curie/l.

The laser-induced decay of Tritium is a competing process that determines the stationary level of Tritium during laser exposure. Note that significant decrease of Tritium content occurs during 1 hour laser exposure, which is many orders of magnitude shorter than life time of this nuclide.

Thus, laser-assisted synthesis of Tritium from $D_2O$ has been realized. The synthesis proceeds under exposure of various solid targets with laser pulses. There is definite threshold of the intensity and duration of laser pulses, the synthesis takes place only with short (10 ps) laser pulses and peak intensities exceeding $10^{11}$ W/cm$^2$. Tritium is synthesized without any cathodic bias on the target though the application of this bias largely increases its rate. Laser exposure of nanoparticles in absence of the solid target does not alter Tritium content.

References

1. Wong, A. Grigoriu, J. Roslund, T.-S. Ho, and H. Rabitz, *Phys. Rev.* A 84, 053429 (2011).

2. Andreev A.V., Gordienko V.M., Dykhne A.M., Saveliev A.B., Tkalya E.V., 1997, *JETP Lett*. **66** (5) 312.

3. Andreev A.V.; Volkov R.V.; Gordienko V.M.; Dykhne A.M.; Mikheev P.M.; Tkalya E.V.; Chutko O.V.; Shashkov A.A.; 1999, *JETP Lett*. 69 (5) 343.

4. Schwoerer H., Gibbon P., Düsterer S., Behrens R., Ziener C., Reich C., Sauerbrey R. *Phys. Rev. Lett.*, 86 (11), 2317 (2001).

5. V. S. Belyaev, A. P. Matafonov, V. I. Vinogradov, V. P. Krainov, V. S. Lisitsa, A. S. Roussetsky, G. N. Ignatyev, V. P. Andrianov, *Phys. Rev. E* 72 (2005), 026406.

6. A.V. Simakin, G.A. Shafeev, arXiv:0906.4268; A.V. Simakin, G.A. Shafeev, *Physics of Wave Phenomena,* 16 (4) (2008) 268-274; A.V. Simakin, G.A. Shafeev, *Journal of Optoelectronics and Advanced Materials*, (2010) 12(3), 432 – 436.

7. A.V. Simakin, G.A. Shafeev, *Quantum Electronics* 41 (7) 614 – 618 (2011).

8. A.V. Simakin, G.A. Shafeev, *Physics of Wave Phenomena* (2011) 19(1), 30–38.

9. G.A. Shafeev, Laser-induced nuclear decays in Uranium isotopes, in: Uranium: Characteristics, Occurrence and Human Exposure, Editors: Alik Ya. Vasiliev and Mikhail Sidorov, ISBN: 978-1-62081-207-5, Novapublishers Inc, New York, 2012, p.p. 117-153.

10. Volkovich A.G., Govorun A.P., Gulyaev A.A. Zhukov S.V., Kuznetsov V.L., Rukhadze A.A., Steblevsky A.V., Urutskoev L.I., Lebedev Bulletin of the Lebedev Physics Institute, 2002, No 8, 45 – 50.

11. G.A. Shafeev, A.V. Simakin, F. Bozon-Verduraz, M. Robert, *Applied Surface Science* **254** (2007) 1022–1026.

12. Fleishmann M., Pons S. and Hawkins M. Electrochemically induced nuclear fusion of deuterium, *J. Electroanal. Chem.* 1989. V. 261. P. 301-308.

13. Iwamura Y., Itoh T., Gotoh N., Toyoda I. Detection of anomalous elements, X-ray, and excess heat in a $D_2$-Pd system and its interpretation by the electron-induced nuclear reaction model, *Fusion Technology*, 1998. V. 33. P. 476-492.

14. Takahashi A. Progress in condensed matter nuclear science. In: Condensed Matter Nuclear Science. Proc. 12[th] Intern. Conf. on Cold Fusion, Yokohamacity, Japan. 27 Nov-2 Dec 2005. World Scientific Publishing Co. Pte. Ltd.: Singapore, 1-25 p.

15. P.V. Kazakevich, A.V. Simakin, V.V. Voronov, and G.A. Shafeev, *Applied Surface Science,* **252** (2006) 4373–4380.

16. P.G. Kuzmin, G.A. Shafeev, V.V. Voronov, R.V. Raspopov, E.A. Arianova, E.N. Trushina, I.V. Gmoshinsky and S.A. Khotimchenko, *Quantum Electronics*, **42**, (2012) 1042.

17. Timashev S.F., Polyakov Yu.S., Lakeev S.G., Misurkin P.I., Danilov A.I. Fundamentals of fluctuation metrology, *Russian Journal Physical Chemistry* A. 2010. V. 84. No 10, 1807-1825.

18. Mirsaidov U., Timashev S.F., Polyakov Yu.S., Misurkin P.I., Polyakov S.V., Musaev I. *Analytical method for parameterizing the random profile components of nanosurfaces imaged by atomic force microscopy* // Analyst. 2011. V. 136. N 3, P. 570-576

19. S. Lau Truong, G. Levi, F. Bozon-Verduraz, A.V. Petrovskaya, A.V. Simakin, G.A. Shafeev, *Applied Surface Science* **254** (2007) 1236–1239.

20. E.V. Barmina, S. Lau Truong, F. Bozon-Verduraz, G. Levi, A.V. Simakin, G.A. Shafeev, , *Quantum Electronics*, 40 (2010) 346-348.

21. Timashev S. F., V. I. Muromtsev V.I., Akovantseva A.A. Nuclear Processes Initiated by Electrons // Russian Journal of Physical Chemistry A, 2013, Vol. 87, No. 6, pp. 1063–1069.

22. Timashev S.F. "Beta-dineutron" as a radioactive element and its possible role in nuclear synthesis processes (in press).

23. Zel'dovich Ya.B., Gol'danskii V.I., Baz' A.I. Systematics of the lightest nuclei, Physics – Uspekhi. Advances in Physical Sciences. 1965. V. 8. N 2. P. 177–201.